\documentclass[a4paper]{jpconf}
\usepackage{graphicx}
\usepackage{amsmath}
\usepackage{amssymb}

\newcommand{\eq}[1]{\begin{align} #1 \end{align}}

\begin{document}
 \title{Updates to the p+p and A+A chemical freeze-out lines from the new experimental data}

\author{\underline{V.~V. Begun}$^1$, V. Vovchenko$^{2,3,4}$, M.~I. Gorenstein$^{2,5}$}

 \address{$^1$ Institute of Physics, Jan Kochanowski University, PL-25406 Kielce, Poland}
 \address{$^2$ Frankfurt Institute for Advanced Studies, Goethe Universit\"{a}t Frankfurt, D-60438 Frankfurt am Main, Germany}
 \address{$^3$ Institut f\"{u}r Theoretische Physik, Goethe Universit\"{a}t Frankfurt, D-60438 Frankfurt am Main, Germany }
 \address{$^4$ Department of Physics, Taras Shevchenko National University of Kiev, 03022 Kiev, Ukraine}
 \address{$^5$ Bogolyubov Institute for Theoretical Physics, 03680 Kiev, Ukraine}

\ead{viktor.begun@gmail.com}

\begin{abstract}
We show that the new data on mean multiplicities measured in p+p
and A+A collisions together with the updated list of resonances
lead to the significant changes of the obtained freeze-out lines.
The new A+A line gives much smaller temperatures at high collision
energies and agrees with the values obtained at the LHC. The newly
obtained p+p line is much closer to the A+A line than previously
expected, and even touches it in the region where the $K^+/\pi^+$
horn appears in the data. It indicates that the temperatures that
will be obtained in the beam energy and system size scan by the
NA61/SHINE Collaboration might be very close. However, our
analysis shows that the chemical potentials could be very
different for the same energies in A+A and p+p. It adds more
puzzles to the set of surprising coincidences at the energies
close to the possible onset of deconfinement.
\end{abstract}

The main motivation of these studies is the new p+p data from the
NA61/SHINE and HADES Collaborations at
$\sqrt{s_{NN}}=3.2-17.3$~GeV~\cite{Abgrall:2013qoa,Agakishiev:2014nim,Pulawski:2015tka,Agakishiev:2015ysr,Aduszkiewicz:2059310},
as well as the new A+A data from HADES, and the updated A+A data
from the NA49 Collaborations at $\sqrt{s_{NN}}=2.2-17.3$~GeV,
see~\cite{Anticic:2011ny,Anticic:2011zr,NA49-latest,Lorenz:2014eja,Vovchenko:2015idt}
and references therein.
A detailed comparison between the description of the p+p and A+A
data at the discussed energies is important, since the ratio of
positively charged kaons to protons, $K^+/\pi^+$, has a sharp
maximum at $\sqrt{s_{NN}}=7.6$~GeV, which was predicted as one of
the signals of the onset of deconfinement in
Ref.~\cite{Gazdzicki:1998vd}, see also the summary of other
deconfinement signals and the latest experimental outcome
in~\cite{Wilczek:2015ava}.

\begin{figure}[h]
\includegraphics[width=0.5\textwidth]{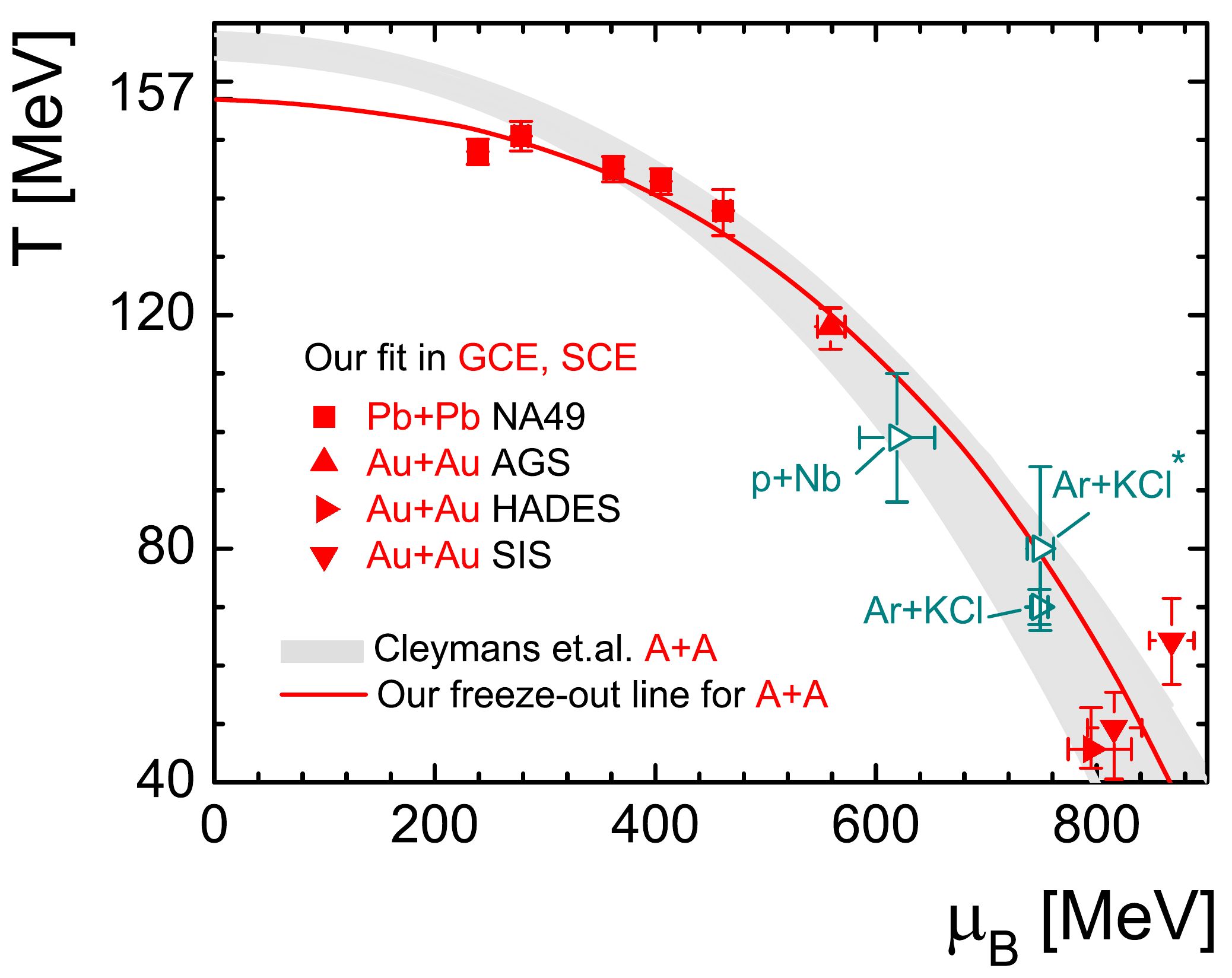}\hspace{0.02\textwidth}%
\begin{minipage}[b]{0.48\textwidth}
\caption{\label{Fig-1} The freeze-out line in central A+A
collisions. The grey band is the previous parametrization
from~\cite{Cleymans:2005xv}. The solid line is our new fit
from~\cite{Vovchenko:2015idt}, see Eqs.~(\ref{Eq-T-mu}),
(\ref{Eq-abc}). The points correspond to different collision
energies. The p+Nb and Ar+KCl points are from the independent
analysis~\cite{Agakishiev:2015bwu}, and were not included in the
fit. The calculations are mostly done in the grand canonical
ensemble (GCE). For the small energies of the old SIS and new
HADES data the exact conservation of strangeness was taken into
account within the strangeness canonical ensemble (SCE)~\cite{Vovchenko:2015idt}. 
}
\end{minipage}
\end{figure}

The updated NA49 A+A data contain more particles, some of them are
with different error bars, while others are excluded. The chemical
freeze-out analysis is performed in the framework of the hadron
resonance gas (HRG) model. We use the latest set of resonances
from THERMUS package~\cite{Wheaton:2004qb} with masses
$M\leq2.4$~GeV, while previous analyzes included only the
resonances with $M\leq1.7$~GeV, see,
e.g.,~\cite{Becattini:2005xt,Cleymans:2005xv}. We also exclude the
$\sigma$ and $\kappa$ resonances from the particle list due to the
reasons described in~\cite{Broniowski:2015oha}. The combination of
these factors alters the freeze-out line obtained within the HRG
see~\cite{Vovchenko:2015idt} and Fig.~\ref{Fig-1}. Our new line in
Fig.~\ref{Fig-1} is the result of the fit to the temperatures $T$
and baryon chemical potentials $\mu_B$ obtained by us at
particular collision energies.
The fit functions for the $T_{\rm A+A}(\mu_B)$ and
$\mu_B(\sqrt{s_{NN}})$ are the same as in~\cite{Cleymans:2005xv},
however, the obtained parameters are rather different:
 \eq{\label{Eq-T-mu}
 &T_{\rm A+A}(\mu_B)~=~ a ~-~ b\mu_B^2 ~-~ c\mu_B^4~,~~~~~~~~~~~~
 \mu_B(\sqrt{s_{NN}})~=~\frac{d}{1~+~e\,\sqrt{s_{NN}}}~,
%
%
\\
\label{Eq-abc}
 &a=0.157\,\text{GeV},~b=0.087\,\text{GeV}^{-1},~c=0.092\,\text{GeV}^{-3},~d=1.477\,\text{GeV},~e=0.343\,\text{GeV}^{-1}.
 }
They suggest that the freeze-out line goes much lower than the
previous estimates from~\cite{Cleymans:2005xv} at small $\mu_B$,
i.e., at large collision energies. For $\mu_B=0$
Eqs.~(\ref{Eq-T-mu}) and (\ref{Eq-abc}) give the temperature
$T\simeq157$~MeV, which was surprisingly obtained in HRG at the
LHC, see discussion in~\cite{Floris:2014pta}. The difference in
the obtained freeze-out parameters is important for the analysis
of heavy nuclei production. Our results agree well with the
LHC~\cite{Stachel:2013zma}, as well as with the NA49
data~\cite{Anticic:2016ckv}.

The new p+p data are much more precise than the previous world
data in that region. It is a difficult test for transport models,
see, e.g., \cite{Vovchenko:2015idt,Vovchenko:2014vda}.
The previous calculations of the chemical freeze-out parameters in
p+p within HRG~\cite{Becattini:1997rv} were performed for larger
collision energies $\sqrt{s_{NN}}\geq17.3$~GeV. The obtained
temperature had a large uncertainty due to the uncertainty in the
existing data at that time. The p+p chemical freeze-out parameters
for $\sqrt{s_{NN}}<17.3$~GeV are calculated for the first time
in~\cite{Vovchenko:2015idt}.

\begin{figure}[h]
\raisebox{-0.50\height}{\includegraphics[angle=0,width=0.46\textwidth]{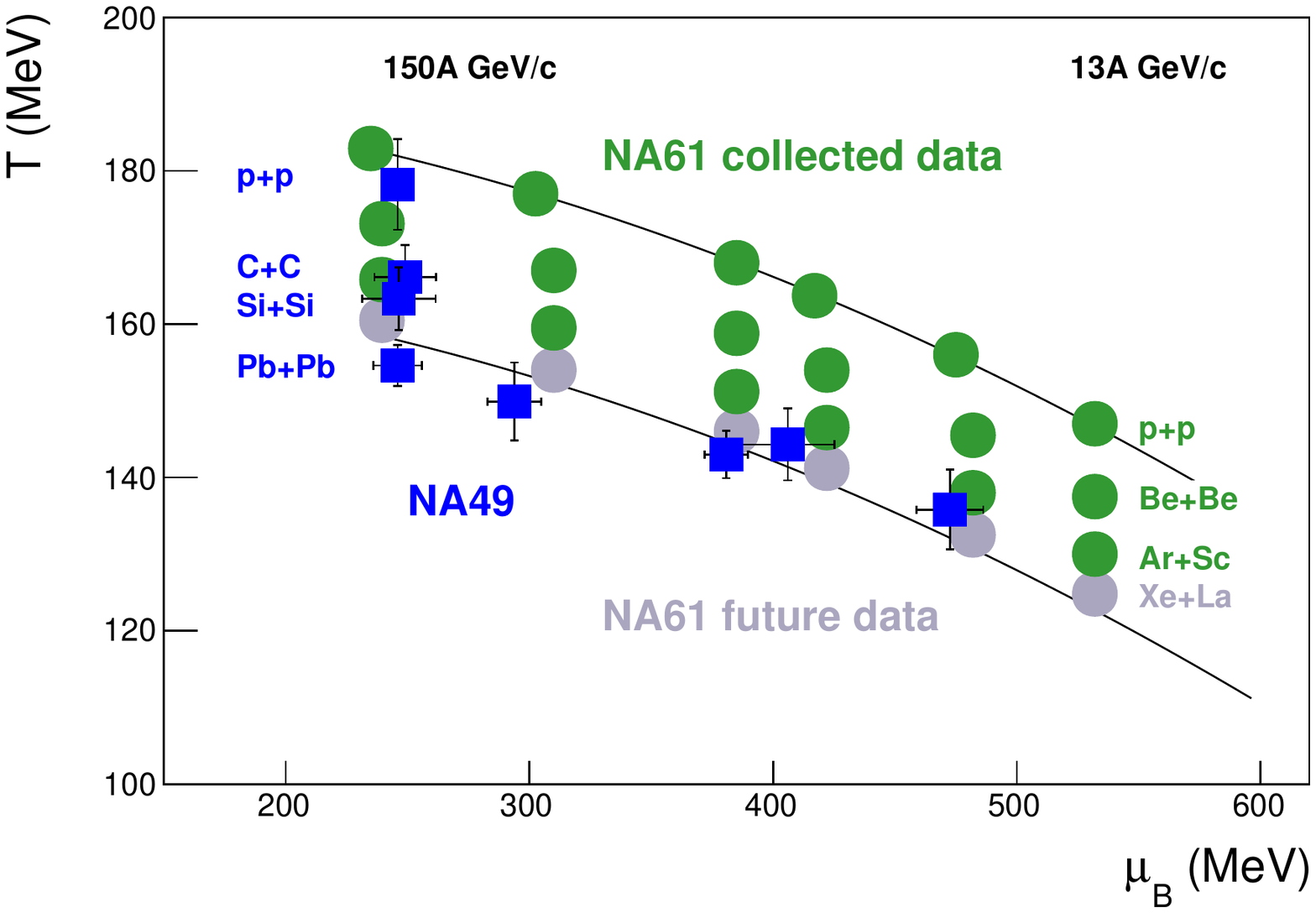}}~~
\raisebox{-0.52\height}{\includegraphics[angle=0,width=0.52\textwidth]{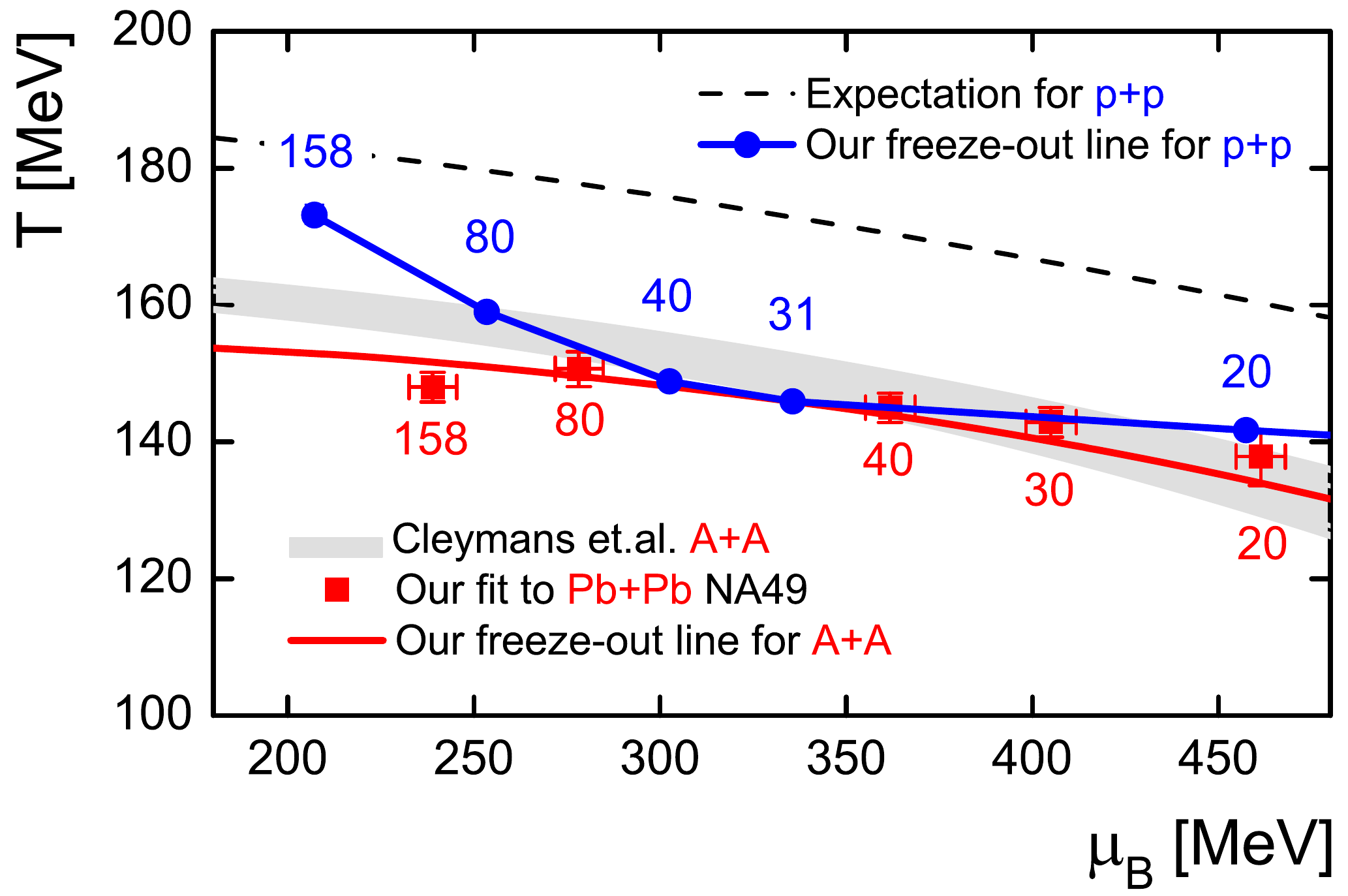}}
\caption{Left: The expectation of the NA61/SHINE Collaboration
(full circles), and the $T(\mu_B)$ values obtained within the HRG
with lighter resonances $M_{\rm res}\leq1.7$~GeV (full squares
with errors) from Ref.~\cite{Gazdzicki:2015ska}. Right: The
$T(\mu_B)$ values obtained for p+p and A+A in our analysis (dots
and solid lines), compared to the expectations in A+A (grey band)
and in p+p (dashed line). The numbers indicate the positions of
the fit results for the corresponding collision energies in the
lab frame $E_{lab}$ in the $A$~GeV units. \label{Fig-2}}
\end{figure}

We performed the fit of the p+p data in the HRG with the latest
table of resonances and excluded $\sigma$, as discussed above. The
analysis is done within the full canonical ensemble (CE) with
exact conservation of electric charge, baryon number, and
strangeness.
Our $T_{\rm p+p}(\mu_B)$ line is obtained in two steps. The
$T_{\rm p+p}(\sqrt{s_{NN}})$ dependance is the straight line fit
to the points obtained within our CE HRG analysis of the new p+p
data~\cite{Vovchenko:2015idt}. The corresponding
$\mu_B(\sqrt{s_{NN}})$ are calculated from the primordial CE HRG
multiplicities of neutrons and anti-neutrons using the relation
between the average baryon number in GCE and the exact baryon
number in CE\footnote{An alternative way is to fit the p+p data
within the GCE, requiring that the obtained temperature is equal
to that in the CE, and, additionally, demand that the average
baryon number, electric charge, and strangeness, equal to the
corresponding exact CE values in p+p, $\langle B\rangle_{\rm
GCE}=B_{\rm CE}=2$, $\langle Q\rangle_{\rm GCE}=Q_{\rm CE}=2$, and
$\langle S\rangle_{\rm GCE}=S_{\rm CE}=0$. This method is
applicable only in thermodynamic limit, i.e. for large enough
systems, but gives practically the same $T_{\rm p+p}(\mu_B)$ line
as in the exact case.}, $\langle B\rangle_{\rm GCE}=B_{\rm CE}$,
see Eqs.~(7-11) in~\cite{Begun:2004zb}. The neutrons and
anti-neutrons are chosen, because they carry only one charge --
the baryon number, and one can use the analytic formulas for the
CE with one charge conservation. The combined result for the
$T(\mu_B)$ in A+A and in p+p is shown in the right panel of
Fig.~\ref{Fig-2}.

One can see that the obtained $T_{\rm p+p}(\mu_B)$ line behaves
very differently, compared to the expectation of the NA61/SHINE in
left panel of Fig.~\ref{Fig-2}. The expected positions of the
$T(\mu_B)$ points for the intermediate systems should be shifted
vertically. They are also much closer to the A+A line than
expected. A similar conclusion can be done looking at the p+Nb and
Ar+KCl points in Fig.~\ref{Fig-1}. It means that the freeze-out
temperatures obtained in the energy and system size scan at the
SPS~\cite{Gazdzicki:2008kk} can be very similar\footnote{One can
still see a clear differences looking at the obtained  radii of
the systems~\cite{Vovchenko:2015idt}.}. The p+p and A+A lines even
touch in the most interesting region $E_{lab}=30-40A$~GeV
($\sqrt{s_{NN}}=7-9$~GeV). However, the chemical potential is
larger in A+A for 70~MeV and for 60~MeV at $E_{lab}=30A$ and
$40A$~GeV, correspondingly, which can be summarized as follows:
 \eq{
 \text{Expectation}&& T_{\rm p+p} &~\gg~T_{\rm A+A}~, && \mu_B^{\rm p+p} ~\simeq~\mu_B^{\rm A+A}~,
 \\
 \text{Our result}&& T_{\rm p+p} &~\simeq~T_{\rm A+A}~, && \mu_B^{\rm p+p} ~\ll~\mu_B^{\rm A+A}~.
 }

The error bars for $T_{\rm p+p}$ are still quite large due to the
small number of measured multiplicities,
see~\cite{Vovchenko:2015idt}. Therefore, more data are needed to
make a firm conclusion.
We found that the minimal set should include particles possessing
all three conserved charges B, S, Q, for both p+p and A+A, for
example, $\pi^{\pm}$, $K^{\pm}$ and $p$, $\bar{p}$ particles.
If the picture seen in right panel of Fig.~\ref{Fig-2} will
preserve after the new measurements, then it will add more puzzles
to the strange coincidences happening at these energies.

\section*{Acknowledgments}
V.V.B. was supported by Polish National Science Center grant No.
DEC-2012/06/A/ST2/00390. M.I.G. was supported by the Program of
Fundamental Research of the National Academy of Sciences of
Ukraine.

\section*{References}
\bibliographystyle{iopart-num}
\bibliography{Begun_SQM2016}

%
\end{document}